# A Novel and Highly Efficient AES Implementation Robust against Differential Power Analysis

Massoud Masoumi

K. N. Toosi University of Tech., Tehran, Iran

m_masoumi@eetd.kntu.ac.ir

## ABSTRACT

Developed by Paul Kocher, Joshua Jaffe, and Benjamin Jun in 1999, Differential Power Analysis (DPA) represents a unique and powerful cryptanalysis technique. Insight into the encryption and decryption behavior of a cryptographic device can be determined by examining its electrical power signature. This paper describes a novel approach for implementation of the AES algorithm which provides a significantly improved strength against differential power analysis with a minimal additional hardware overhead. Our method is based on randomization in composite field arithmetic which entails an area penalty of only 7% while does not decrease the working frequency, does not alter the algorithm and keeps perfect compatibility with the published standard. The efficiency of the proposed technique was verified by practical results obtained from real implementation on a Xilinx Spartan-II FPGA.

## 1. INTRODUCTION

Since the first publication of Kocher et al. [1] titled "Differential Power Analysis", many power analysis attacks have been developed. Power analysis attacks exploit the dependence between the instantaneous power consumption of a cryptographic device and the data it processes and/or the operation it performs. This type of attacks is known as a powerful technique for revealing confidential data (e.g. a secret key of a cryptographic algorithm) because it is imperceptible to users and it does not require expensive equipments like the micro probing techniques [2, 3]. The Differential Power Analysis (DPA) is a statistical test which examines a large number of power consumption signals to retrieve secret keys. The DPA is multiform itself. It can be performed by analyzing the intermediate values of one bit (mono-bit DPA) or a set of several bits (multi-bit DPA). It can also be observed at one instant of time [1] (first-order DPA) or at some instants of time (higher-order DPA) [4]. Another form of these attacks, the so called Correlation Power Analysis (CPA) technique based on the correlation between the real power consumption of the device and a power consumption model, has been widely studied in the literature [5]. In recent years, the security of the Advanced Encryption Standard (AES) against DPA has received considerable attention and there is a growing interest in efficient and secure realization of the AES. As a result of these attacks, numerous hardware and algorithmic countermeasures have been proposed. Unfortunately, most of these techniques are inefficient or costly or vulnerable to higher-order attacks [6]. They include randomized clocks, memory encryption/decryption schemes [7], power consumption randomization [8], and decorrelating the external power supply from the internal power consumed by the chip. Moreover, the use of different hardware logic, such as complementary logic, sense amplifier based logic (SABL), and asynchronous logic [9, 10] have been also proposed. Some of these techniques require about twice as much area and will consume twice as much power as an implementation that is not protected against power attacks. For example, the technique proposed in [10] adds area 3 times and reduces throughput by a factor of 4. Another method is masking which involves ensuring the attacker cannot predict any full registers in the system without making run-specific assumptions that are independent of the inputs to the system. This is achieved by applying a reversible random mask to the plaintext data before encryption with a modified algorithm. This makes exploiting data from several encryptions impossible as it would require guessing the correct mask for each run. Unfortunately, this method is costly or inefficient even if it has been demonstrated that it works. The main problem with masking methods is that they usually require an extra datapath that work in parallel to compute the modification of the mask by the algorithm which considerably increases hardware overhead and decreases the throughput. Most importantly, some masking techniques that were proposed were shown to be susceptible to higher order DPA attacks. Even techniques that were shown to be theoretically provably secure were susceptible to DPA using predictions based on simulations and a back-annotated netlist [11].

In this work, we concentrate on algorithmic countermeasures to protect AES against power attack and present a novel core implementation which is very simple and effective with very low hardware cost. This countermeasure is based on mathematical properties of Rijndael algorithm, and retains perfect compatibility with the published standard. We have studied the use of composite field techniques and isomorphism for Galois Field arithmetic in the context of protection of Rijndael against differential power attack. In order to experimentally verify the effectiveness of our proposed countermeasure we have implemented two versions of AES, a protected and an unprotected, on a Xilinx Spartan-II FPGA and compared the results of the implementation in the terms of resistance against attack, speed, area and throughput. While FPGAs are becoming increasingly popular for cryptographic applications, there are only a few articles that assess their vulnerability to such attacks. In particular, very little work has been done on the resistance of FPGAs to hardware or system attacks, which, in practice, pose far a greater danger than algorithmic attacks. The results we obtained in this work are very encouraging compared with the results reported in the literature. The area overhead is only 7% with no decrease in speed or clock frequency or alteration in the algorithm. In addition, our technique directed at both hardware and software realizations and could be easily used in variety of platforms such as FPGAs, smart cards, DSPs or other security tokens. Most importantly, this work shows that it is possible to design algorithms to be inherently impervious to DPA. This article is organized as follows: The AES algorithm will be briefly discussed in section II. The principle of DPA attack will be reviewed in section III. The proposed approach will be illustrated in section VI. Section V explains the measurement setup used for the implementation of the attack and the results obtained. Finally, we summarize the results of our work in the conclusions.

## 2. THE AES ALGORITHM

AES has been developed and published by Daemen and Rijmen [12]. This algorithm is a byte-oriented symmetric block cipher, composed of a sequence of four primitive functions, Sub Bytes, ShiftRows, MixColumns, and AddRoundKey, executed round by round. Prior to each round AddRoundKey which combines the input with the cipher key is executed. In a 128-bit operation mode, at the start of the encryption, the message is divided to the blocks of length 128-bit and is copied to a 16 byte rectangular array called State. AddRoundKey is only a simple bit-wise XOR operation in which the elements of the State are XORed with RoundKey bit-by-bit. Sub Bytes is a non-linear bit-wise substitution of all bytes in the State. In Sub Bytes, each byte in the State is replaced by its corresponding byte in another table called S-Box. S-Box contains multiplicative inverse of all possible bytes over $GF(2^8)$ followed by an affine transformation. Each byte is an element of Galois field $GF(2^8)$ with irreducible polynomial $m(x) = x^8 + x^4 + x^3 + x + 1$. In the ShiftRows transformation, each row of the state is considered separately and the bytes in that row are cyclically shifted to the left based upon the key-size of the algorithm. For the 128-bit key, the first row is unchanged. However, the second, third and fourth rows are shifted one, two, and three bytes respectively. The MixColumns transformation is a bricklayer permutation operating on each column of the State. In MixColumns, columns of the State are considered as a four-term polynomial over $GF(2^8)$, then are multiplied with a fixed polynomial $c(x) = \{03\}x^3 + \{01\}x^2 + \{01\}x + \{02\}$. Multiplications are performed modulo $(x^4+1)$. The algorithm for the decryption has the same structure but uses mathematical inverses of the encryption steps, i.e. InvSubBytes, InvShiftRows, and InvMixColumns. The round keys are the same as those in encryption but are used in reverse order. Figure 1 shows the standard implementation of the AES.

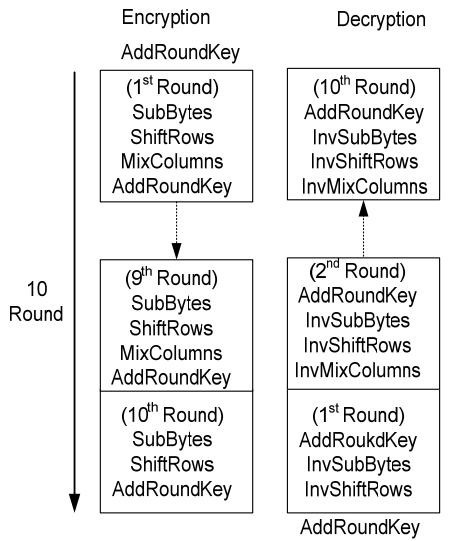

**Figure 1. Standard implementation of the AES algorithm.**

## 3. DPA AGAINST AES

In DPA, an attacker uses a so-called hypothetical model of the attacked device. The model is used to predict several values for the side-channel output of a device. This hypothetical model for the AES is one AddRoundKey and the SBox lookup of the first round which is fed with the plaintexts and one byte of the first subkey. The output of SubBytes is usually attacked in practice since that is the only function in AES in which data and cipher key enter a direct operation. These predictions are compared to the real, measured side-channel output of the device. Comparisons are performed by applying statistical methods on the data. Among others, the most popular are the *distance-of-mean test* and the *correlation analysis*. For the correlation analysis, the model predicts the amount of side-channel leakage for a certain moment of time in the execution. These predictions are correlated to the real side-channel output. This correlation can be measured using the Pearson correlation coefficient. Let $t_i$ denotes the $i^{th}$ measurement data (i.e. the $i^{th}$ trace) and $T$ the set of traces. Let $p_i$ denote the prediction of the model for the $i^{th}$ trace and $P$ the set of such predictions [13]. Then we calculate:

$$C(T,P) = \frac{E(T.P) - E(T)E(P)}{\sqrt{Var(T).Var(P)}} \quad (1)$$

Here $E(T)$ denotes the expectation (average) trace of the set of traces $T$ and $Var(T)$ denotes the variance of a set of traces $T$. In practice it is not possible to know the true values for the covariance or standard deviation of variables, only calculate approximations of them based on the values discovered through experiments. If this correlation is high, it is usually assumed that the prediction of the model, and thus the key hypothesis, is correct. The scenario of DPA attack based-on distance-of-mean test is as follows. At first $N$ plaintexts are randomly generated. Power consumption measures are taken for each plaintext. As before, a hypothetical model of the AES is fed with the plaintexts and one byte of the first subkey. Only the SBox of the first round is targeted by the attacker since that is the only function in AES in which data and cipher key enter a direct operation (See Figure 1). To this output hypothesis, a selection function *D* is applied. This selection function divides the measures in two sets. One where the selection function returns one and the other for the return zero. For each set the average is computed. Then, the difference between the two averages is calculated. This leads to $2^8$ differential curves. Only for the correct subkey the selection function has worked properly and there will be well seen spikes in an otherwise flat curve [1]. High-order DPA uses more general DPA selection functions to perform differential power analysis. High-order attacks require using multiple samples in a single power trace to compute a DPA power trace value. Using multiple samples is analogous to second order or higher digital signal processing with memory. An attacker can mount a second order attack by computing joint statistics on power signatures at different sections on the encryption code. One drawback to high-order DPA is increased memory and processor requirements because of the need to store multiple samples for a single DPA computation. The required trace equipment for high-order DPA is identical to the trace equipment required for SPA and DPA, but more sophisticated post-processing requires additional off-line resources. Knowledge of the encryption algorithm and specific implementation is more critical in high-order DPA than first-order. Attackers need to know specific points of execution where joint statistics can be meaningfully computed.

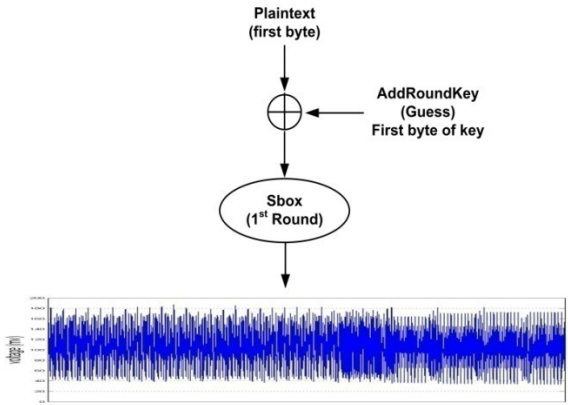

**Figure 1. Partial power trace of an AES encryption.**

## 4. AES AND COMPOSITE FIELD ARITHMETIC

The SubBytes and the InvSubBytes in the AES algorithm are traditionally implemented by look-up tables (LUT). Non-LUT-based approaches, which employ combinational logic only, such as the composite field (or tower field) inversion over $GF(2^8)$ are used to avoid the unbreakable delay of LUTs, and it can be used to create compact AES implementations [13]. Composite field arithmetic can be employed, such that the field elements of $GF(2^8)$ are mapped to elements in some isomorphic composite fields, in which the field operations can be implemented by lower cost subfield operations. The two pairs $\{GF(2^n), Q(y)\}$ and $\{GF(2^n)^m), P(x)\}$ constitute a composite field if $GF(2^n)$ is constructed from $GF(2)$ by $Q(y)$ and $GF((2^n)^m)$ is constructed from $GF(2^n)$ by $P(x)$, where $Q(y)$ and $p(x)$ are polynomials of degree n and m respectively. The fields $GF((2^n)^m)$ and $GF(2^k)$, k = nm, are isomorphic to each other. The most costly operation in the SubBytes is the multiplicative inversion over a field A (the AES field), where A is extended from GF(2) with the irreducible polynomial m(x) mentioned in Section II. To reduce the cost of this operation, the following 3-stage method is adopted:
(Stage 1) Map all elements of the field A to a composite field B, using an isomorphism function δ.
(Stage 2) Compute the multiplicative inverses over the field B.
(Stage 3) Re-map the computation results to A, using the function $δ^{-1}$. Figure 2 shows the outline of an S-Box implementation using the composite field technique.

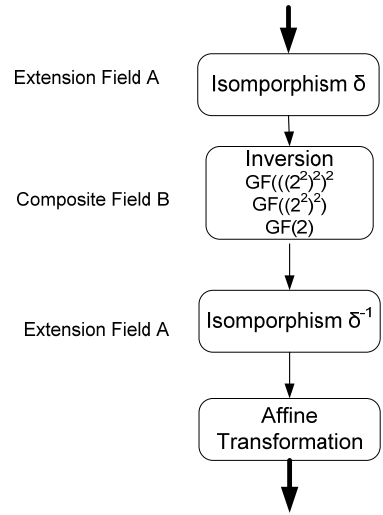

**Figure 2. Computation sequences of composite-field based SBox.**

To reduce the cost of Stage 2 as much as possible, it is known to be efficient to construct the composite field B using repeated degree-2 extensions under a polynomial basis using these irreducible polynomials [13],

$$\begin{cases} GF(2^2) & \to P_0(x) = x^2 + x + 1 \\ GF(((2^2)^2)) & \to P_1(x) = x^2 + x + \phi \\ GF(((2^2)^2)^2) & \to P_2(x) = x^2 + x + \lambda \end{cases} \quad (2)$$

Where $\Phi = \{10\}_2$, $\lambda = \{1100\}_2$. The isomorphism functions δ and $δ^{-1}$ in Stages 1 and 3 are constructed as follows. The δ (and $δ^{-1}$) can be found as follows.

$$\delta = \begin{bmatrix} 1 & 0 & 1 & 0 & 0 & 0 & 0 & 0 \\ 1 & 1 & 0 & 1 & 1 & 1 & 1 & 0 \\ 1 & 0 & 1 & 0 & 1 & 1 & 0 & 0 \\ 1 & 0 & 1 & 0 & 1 & 1 & 1 & 0 \\ 1 & 1 & 0 & 0 & 0 & 1 & 1 & 0 \\ 1 & 0 & 0 & 1 & 1 & 1 & 1 & 0 \\ 0 & 1 & 0 & 1 & 0 & 0 & 1 & 0 \\ 0 & 1 & 0 & 0 & 0 & 0 & 1 & 1 \end{bmatrix}$$

$$\delta^{-1} = \begin{bmatrix} 1 & 1 & 1 & 0 & 0 & 0 & 1 & 0 \\ 0 & 1 & 0 & 0 & 0 & 1 & 0 & 0 \\ 0 & 1 & 1 & 0 & 0 & 0 & 1 & 0 \\ 0 & 1 & 1 & 1 & 0 & 1 & 1 & 0 \\ 0 & 0 & 1 & 1 & 1 & 1 & 1 & 0 \\ 1 & 0 & 0 & 1 & 1 & 1 & 1 & 0 \\ 0 & 0 & 1 & 1 & 0 & 0 & 0 & 0 \\ 0 & 1 & 1 & 1 & 0 & 1 & 0 & 1 \end{bmatrix}$$

Let $q$ be the element in $GF(2^8)$, then the isomorphic mappings and its inverse can be written as $\delta*q$ and $\delta^{-1}*q$, which is a case of matrix multiplication as shown below, where $q_7$ is the most significant bit and $q_0$ is the least significant bit. The matrix multiplication can be translated to logical XOR operation as is shown below.

$$\delta \times q = \begin{bmatrix} q_7 \oplus q_5 \\ q_7 \oplus q_6 \oplus q_4 \oplus q_3 \oplus q_2 \oplus q_1 \\ q_7 \oplus q_5 \oplus q_3 \oplus q_2 \\ q_7 \oplus q_5 \oplus q_3 \oplus q_2 \oplus q_1 \\ q_7 \oplus q_6 \oplus q_2 \oplus q_1 \\ q_7 \oplus q_4 \oplus q_3 \oplus q_2 \oplus q_1 \\ q_6 \oplus q_4 \oplus q_1 \\ q_6 \oplus q_1 \oplus q_0 \end{bmatrix}$$

$$\delta^{-1} \times q = \begin{bmatrix} q_7 \oplus q_6 \oplus q_5 \oplus q_1 \\ q_6 \oplus q_2 \\ q_6 \oplus q_5 \oplus q_1 \\ q_6 \oplus q_5 \oplus q_4 \oplus q_2 \oplus q_1 \\ q_5 \oplus q_4 \oplus q_3 \oplus q_2 \oplus q_1 \\ q_7 \oplus q_4 \oplus q_3 \oplus q_2 \oplus q_1 \\ q_5 \oplus q_4 \\ q_6 \oplus q_5 \oplus q_4 \oplus q_2 \oplus q_0 \end{bmatrix}$$

Thus, the multiplicative inversion in $GF(2^8)$ can be carried out in $GF((2^4)^2)$ by the architecture illustrated in Figure 3.
The multipliers in $GF(2^4)$ can be further decomposed into multipliers in $GF(2^2)$ and then to $GF(2)$, in which a multiplication is simply an AND operation. Figure 3 illustrates this decomposition, together with the other blocks used in Figure 4 except the inversion in $GF(2^4)$ block.

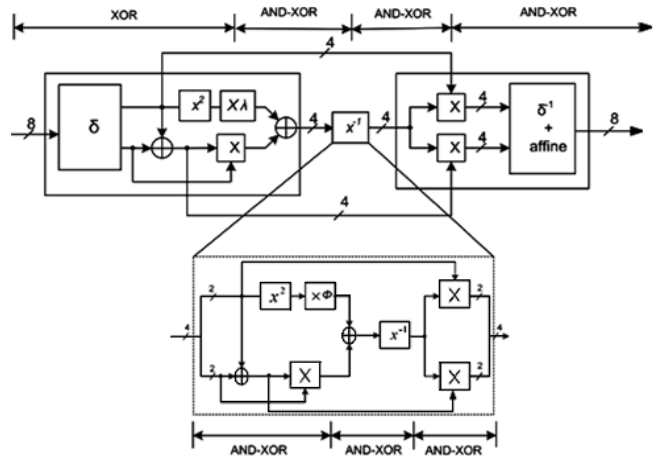

**Figure 3**. **Implementation of SubBytes Transformation using composite field arithmetic.**

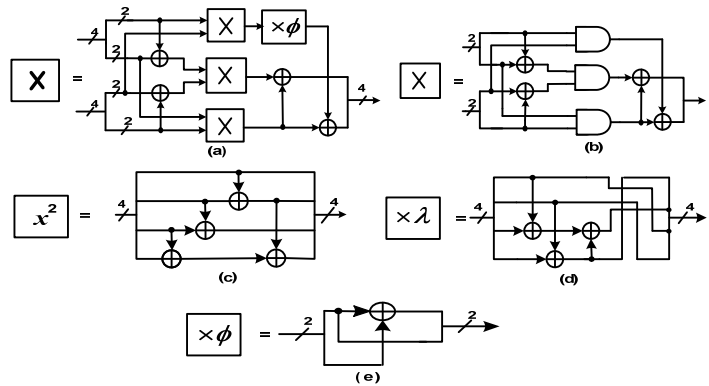

**Figure 4. Implementation of individual blocks, (a) multiplier in $GF(2^4)$, (b) multiplier in $GF(2^2)$, (c) squarer in $GF(2^4)$, (d) constant multiplier (×λ); and (e) multiplier(×Φ).**

## 3.1 Composite Field Arithmetic Operations

Any arbitrary polynomial can be represented by $bx + c$ where $b$ is upper half term and $c$ is the lower half term. Therefore, from here, a binary number in Galois Field $q$ can be spilt to $q_H x + q_L$. For instance, if $q = \{1011\}_2$, it can be represented as $\{10\}_2 x + \{11\}_2$, where $q_H$ is $\{10\}_2$ and $q_L = \{11\}_2$. $q_H$ and $q_L$ can be further decomposed to $\{1\}_2 x + \{0\}_2$ and $\{1\}_2 x + \{1\}_2$ respectively. The decomposing is done by making use of the irreducible polynomials introduced at (2). Using this idea, the logical equations for the addition, squaring, multiplication and inversion which were shown in Figure 4 can be derived. Detailed explanation of the implementation of all these blocks is out of the scope of this article. However, to clarify the subject to the readers to better understand how these modules can be implemented using combinational logic, the realization of constant multipliers ($\times \lambda$) and ($\times \varphi$) is briefly illustrated [14]. For the purpose of practicality, the depth of the mathematics involved has been reduced in order to allow the reader to better figure out the internal operations within the S-Box.

Let $k = q\lambda$, where $k = \{k_3\ k_2\ k_1\ k_0\}_2$, $q = \{q_3\ q_2\ q_1\ q_0\}_2$ and $\lambda = \{1100\}_2$ are elements of $GF(2^4)$.

$k = \{k_3 k_2 k_1 k_0\} = k_H x + k_L = \{q_3 q_2 q_1 q_0\}(1100)$
Where $\lambda_H = \{11\}_2$ and $\lambda_L = \{00\}_2$

$k = (q_H x + q_l)(\lambda_H x + \lambda_l)$
$\lambda_l$ can be canceled out since $\lambda_l = \{00\}_2$
$k = q_H \lambda_H x^2 + q_l \lambda_H x$

Modulo reduction can be performed by substituting $x^2 = x + \varphi$ using the irreducible polynomial in (2) to yield the expression below.

$k = q_H \lambda_H (x+\varphi) + q_L \lambda_H x$
$k = (q_H \lambda_H + q_l \lambda_H)x + (q_H \lambda_H)\ \varphi \in GF(2^2)$

$k_H$ and $k_L$ terms can be further broken down to GF(2).

$k_H = q_H \lambda_H + q_L \lambda_H$
$k_H = (q_3 q_2)(11_2) + (q_1 q_0)(11_2)$
$k_H = (q_3 x + q_2)(x+1) + (q_1 x + q_0)(x+1)$
$k_H = q_3 x^2 + (q_3 + q_2)x + q_2 + q_1 x^2 + (q_1 + q_0)x + q_0$ (3)

Substituting $x^2 = x + 1$, would then yield the following.

$k_H = q_3(x+1) + (q_3+q_2)x + q_2 + q_1(x+1) + (q_1+q_0)x + q_0$
$k_H = (q_3+q_3+q_2+q_1+q_1+q_0)x + (q_3+q_2+q_1+q_0)$
$k_3 x + k_2 = (q_2+q_0)x + (q_3+q_2+q_1+q_0) \in GF(2)$

The same procedure is taken to decompose $k_L$ to GF(2).

$k_L = q_H \lambda_H \varphi$
$k_L = (q_3 q_2)(11_2)(10_2)$
$k_L = (q_3 x + q_2)(x+1)x$
$k_L = q_3 x^3 + q_2 x^2 + q_1 x + q_0$

Again, the $x^2$ term can be substituted since $x^2 = x + 1$. Likewise, $x^3$ is also substituted with $x^3 = 1$. So, we have:

$k_L = q_3(1) + q_2(x+1) + q_3(x+1) + q_2 x$
$k_L = (q_3+q_2+q_1)x + (q_3+q_2+q_1)$
$k_1 x + k_0 = (q_3)x + (q_2) \in GF(2)$ (4)

From equations (3), (4) combined, the formula for computing multiplication with constant $\lambda$ is shown below.

$k_3 = q_2 \oplus q_0$
$k_2 = q_3 \oplus q_2 \oplus q_1 \oplus q_0$
$k_1 = q_3$
$k_0 = q_2$ (5)

From the equation (5) the architecture for multiplication with constant $\lambda$ can be depicted as below ($\Phi = \{10\}_2, \lambda = \{1100\}_2$).

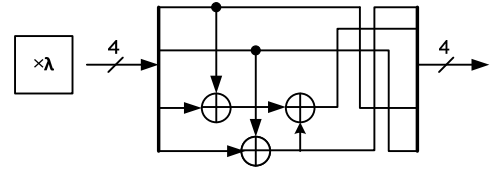

**Figure 5. Hardware diagram for multiplication with constant $\lambda$ ($\lambda = 12$)**

The same method could be used for multiplication with constant $\lambda$ when $\lambda = \{1111\}_2$. Figure 6 shows this architecture. Other architectures shown in figure 4 can be implemented simply using the same technique. As it is seen any change in $\{\Phi, \lambda, \delta, \delta^{-1}\}$ could be easily translated to Boolean logic and changing the SBox architecture by using this method has a very low hardware cost.

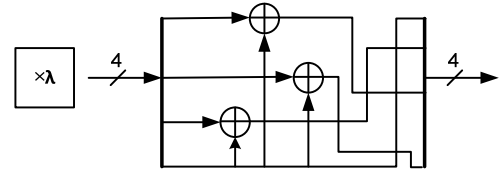

**Figure 6. Hardware diagram for multiplication with constant $\lambda$ ($\lambda = 15$)**

As another example consider multiplication with constant ($\times \varphi$) in $GF(2^2)$ where $\varphi = \{10\}_2$. Let $k = q\varphi$ where $k = \{k_1 k_0\}_2$, $q = \{q_1 q_0\}_2$ and $\varphi = \{10\}_2$ are elements of $GF(2^2)$.

$k = k_1 x + k_0 = (q_1 q_0)(10_2) = (q_1 x + q_0)(x)$
$k = q_1 x^2 + q_0 x$
Substitute the term $x^2$ with $x+1$, yield the expression below.
$k = q_1(x+1) + q_0 x$
$k = (q_1 + q_0)x + (q_1) \in GF(2)$
hence, the formula for computing multiplication with $\varphi$ can be derived and is shown below.
$k_1 = q_1 + q_0$
$k_0 = q_1$
The hardware implementation of multiplication with $\varphi = \{10\}_2$ is shown in figure 7.

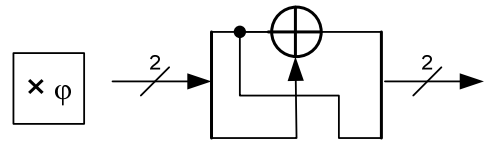

**Figure 7. Hardware implementation of multiplication with $\varphi = \{10\}_2$.**

The hardware implementation of the multiplication with $\varphi = \{11\}_2$ can be obtained as follows.
$k = k_1 x + k_0 = (q_1 q_0)(11_2) = (q_1 x + q_0)(x+1)$
$k = q_1 x^2 + (q_0 + q_1)x + q_0$
Substitute the term $x^2$ with $x+1$, yield the expression below.
$k = q_0 x + (q_0+q_1) \in GF(2)$
hence, the formula for computing multiplication with $\varphi = \{11\}_2$ can be derived and is shown in figure 8.
$k_1 = q_0$
$k_0 = q_0 + q_1$

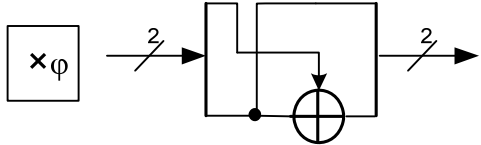

**Figure 8. Hardware implementation of multiplication with $\varphi = \{11\}_2$.**

## 3.2 PREVIOUS WORKS

In [15] Rostovtsev and Shemyakina describe using of isomorphisms of the underlying finite field. But as the authors admit, their method has comparatively small efficiency. The technique proposed in [16] randomizes power consumption of SBox by randomly choosing irreducible generator polynomials of the field $GF(2^8)$. The approach is interesting but introduces significant hardware cost with almost 300% increase in area and 60% decrease in speed. Another method based on univariate polynomials of Blomer et. al. has been illustrated in [17]. This can be seen as a perfectly general method that can be applied to any S-box, as any function over a finite field can be seen as a univariate polynomial. This makes the method more efficient than in the general case and suitable for Rijndael. However, it suffers from considerable decrease in performance. Tower Fields Methods by Oswald, et al. [18, 19] are designed for hardware implementations. In these methods, the computing of Inv in $GF(2^{2k})$ is reduced to a secure computation with masked values of multiplications and inverses in $GF(2^k)$, by representing $GF(2^{2k})$ as a quadratic extension of $GF(2^k)$. Multiplications can be computed with additive masking and we are left with the problem of a secure computation of Inv at the lower level. In [20] Schramm proposes mask multipliers for $GF(2^2)$ and $GF(2^4)$ which are used in the masked composite field-based AES SBox for software applications. However, this approach needs 1536 bytes ROM to store the masked SBox and takes 13600 clock cycles to implement masked AES encryption while a similar unmasked AES realization takes 800 cycles.

## 4 NEW IDEA

In the power analysis, the key detection is possible because of the dependency between the power consumption of devices and the intermediate values of the cryptographic algorithms. Therefore, if we want to prevent from these kinds of attacks, this dependency should be broken. Considering this fact, we have studied the structure of SBox and based on our knowledge of math, the three parameters $\{\Phi, \lambda, \delta\}$ are not constant since there could be much more than one eligible isomorphism (see Figures 3, 4). The polynomial $P_1(x) = x^2 + x + \phi$ is an irreducible polynomial over the field $GF(2^2)$ if and only if $1 < \Phi < 4$, and $P_2(x) = x^2 + x + \lambda$ is irreducible over $GF(2^4)$ if and only if $7 < \lambda < 16$. So, there are 16 different combinations for $\{\phi, \lambda\}$.

We can also search in the set of linear transformations from $GF(2^8)$ to itself or all autoisomorphism over $GF(2^8)$ and it can be proved that there are $\prod_{i=0}^{7}(2^8 - 2^i)$ of such transformations. So we can have the same number of different $\delta/\delta^{-1}$ [21, 22] and considerable number of sets $\{\Phi, \lambda, \delta, \delta^{-1}\}$ for Rijndael. It must be mentioned that all of the possible combinations of $\{\Phi, \lambda, \delta, \delta^{-1}\}$ will not result in an appropriate field isomorphism. We have found 32 different suitable sets of the mentioned parameters which can be implemented with minimum combinational logic. For example, these three following sets have such characteristics. The elements of matrices $\delta$ and $\delta^{-1}$ and the values of $\Phi$ and $\lambda$ are represented in decimal.

$\Phi = 2, \lambda = 15$
$\delta = \{160, 126, 114, 162, 182, 84, 16, 217\}^T$
$\delta^{-1} = \{46, 28, 174, 2, 122, 26, 144, 75\}^T$
$\Phi = 3, \lambda = 12$
$\delta = \{160, 222, 172, 174, 202, 238, 44, 227\}^T$
$\delta^{-1} = \{102, 212, 230, 162, 10, 234, 176, 233\}^T$
$\Phi = 3, \lambda = 10$
$\delta = \{160, 126, 172, 2, 20, 132, 130, 99\}^T$
$\delta^{-1} = \{190, 132, 62, 106, 98, 2, 112, 141\}^T$

Therefore, one can make a random isomorphism by generating different sets of $\{\Phi, \lambda, \delta, \delta^{-1}\}$ and randomly choosing them in each block encryption/decryption. As described, two ciphers are isomorphic if they produce the same output for the same input. Hence, from the architectures shown in Figure 3 and Figure 4, it is obvious that we can have many SBoxes with similar input/output pairs and different internal structures. As it was shown in section III, it is simple to realize and select different architectures of SBox based on different values of $\{\Phi, \lambda, \delta, \delta^{-1}\}$ since such changes can be easily realized by Boolean gates only (please see figures 4-6). As a result, for a specific input, there will be different power consumption patterns while the output remains the same. Since different binary values are used each time, power traces collected for DPA will be weakly correlated to the data being manipulated and it will be harder to extract the secret key or any other sensitive information, even if many runs are performed for the same inputs. The proposed technique effectively reduces the signal to noise ratio (SNR) of the performed operations. As we know, the SNR quantifies how much information is leaking from a power trace. The most important features of the proposed method are that it does not change the mathematical properties of SBox at all, do not decrease the working frequency and has very low area overhead because any change in the architecture of Figure 3 and Figure 4 can be easily realized by Boolean gates only. In addition, it is very simple compared with those presented in the literature and fully complies with the published standard. No additional parameters than the secret key and the data to be processed are needed. The operation of the system is as follows: at first all 32 (or more) possible sets of values for $\{\Phi, \lambda, \delta, \delta^{-1}\}$ are produced and stored. In our prototype, a linear feedback shift register (LFSR) is used as a pseudorandom number generator for proof of the concept. Encryptor and decryptor agree on the initial state of this LFSR as they agree on the cipher key. In each block encryption/decryption, one set is chosen by pseudorandom number generator and one of the 32 corresponding SBox architectures (as was shown in figure 4) is selected randomly. Please notice that multiplication by different values of $\{\Phi, \lambda, \delta\}$ is sufficient to change the intermediate values of these 32 Sboxes and their other parts are common and do not change (e.g., squarer, multiplier in $GF(2^2)$, …). This is why this approach has very low hardware cost. This results in very low area overhead with no decrease in speed or clock frequency since no hardware module has been added to the critical path. In order to prevent any power information leakage when the output of SBox is stored in the output register, another random number generator selects two other $\delta^{-1}$ matrixes randomly and multiplies the pre-final value to them concurrently in a parallel path. This make power consumption pattern of the final stage of SBox computation indistinguishable for the attacker and to prevent he/she to mount a successful attack on this point. This concurrent multiplication can be easily performed in FPGA since it is one of the inherent advantages of FPGA. The simplicity of this approach makes it suitable for both hardware and software applications. Since the mentioned parameters could be easily stored inside the device, the proposed approach could be used in smart cards, digital signal processors or other security tokens.

## 5. ATTACK ON A REAL SYSTEM

To investigate the effect on efficiency of implementing a device using the new style, both regular AES and the new design were realized in a loop architecture in which only one block of data is being processed at a time, using Verilog HDL and were synthesized on a Xilinx Spartan-II FPGA. Both used 128-bit keys. The encryptor core structure in regular AES occupied 983 CLB slices (41%), on targeting Xilinx Spartan-II FPGA device while this number for the modified implementation with 32 randomly selected SBoxs was 1050. This means that the total area cost for the implementation is almost 7%, much lower than the results reported in the literature so far. The clock speed in both cases was almost 42.2 MHz. Our measurement setup, as it is shown in Figure 9, consists of the FPGA board, an Agilent MSO7034A sampling oscilloscope with a 2 GS/sec, BW = 350 MHz. The board uses two separate power supplies, a 3.3 V supply for I/O and a 2.5 V supply for the core cells. Only the core power supply was measured. A small resistor (10 ohm) was inserted between the FPGA board and the power supply. In order to reduce switching noise and to improve the accuracy, the working frequency of FPGA board was lowered to 1 KHz and all measurements were averaged over ten times.

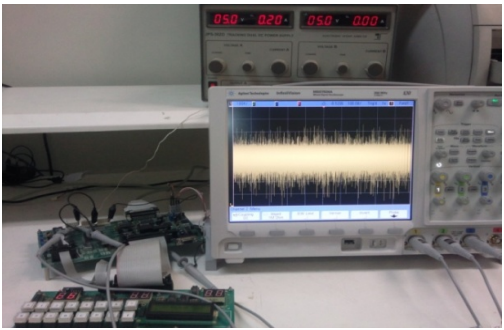

**Figure 9. Experimental setup used for mounting the attack.**

In order to improve the SNR and accuracy of the attack, instead of considering a monobit, a set of four bit was considered, i.e. our selection function returns one when hamming weight of the output of SBox is greater than four, and otherwise it returns zero. It has been shown that the efficiency of the attack is increased in such a case since the ghost peaks and secondary peaks are lowered [23].

The experimental results for the differential power traces for the correct and a wrong subkey guesses are shown in Figure 10 and Figure 11 respectively. As it is seen, the plots confirm the assumption about the measurability of Hamming-Weights leakage. We need approximately 1,000 measurements to identify the correct key.

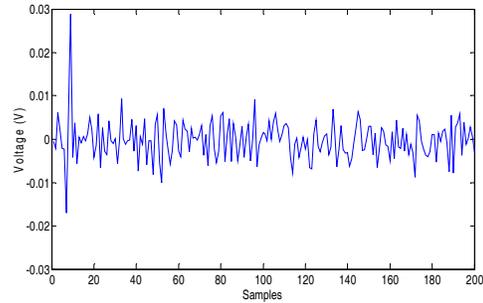

**Figure 10. The differential power traces for the correct subkey guess in the unprotected implementation.**

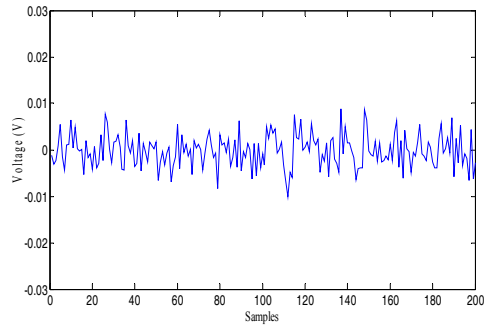

**Figure 11. The differential power traces for a wrong subkey guess in the unprotected implementation.**

The results of calculation of correlations for the first subkey are shown in Figure 12, in which the correct value, 0x3C, appears as a clear peak. Our experiments showed that recovering the full 128-bit key with this method takes almost two hours with an Intel 2.5 GHz quad processor computer.

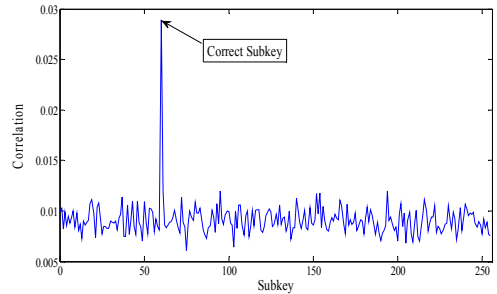

**Figure 12. Recovering the correct subkey using a correlation attack with real measurements.**

The setup was repeated using an implementation of the new algorithm. Figure 13 and Figure 14 show the differential power traces for the correct and wrong key guesses in the protected version respectively. As it is seen the correct key cannot be distinguished from the wrong key as there is more much noise in the system. Figure 15 and Figure 16 show the different patterns of power traces when inputting the same plaintext. Figure 17 shows the correlations for each possible key guess in the new implementation after 1000 traces. This time the correct value is completely obscured by other values and there are no clear peaks, just a band of random values. The experiment was repeated with a new key schedule, this time 6000 power traces were recorded. Again, the correct value was not distinguishable from the incorrect ones. This protection technique could be combined with other randomization techniques to make the probability of success of DPA as low as possible. For example, the technique mentioned above could be used with randomly inserting dummy codes and/or shuffling. As we know, the basic idea of shuffling is to randomly change the sequence of those operations of a cryptographic algorithm that can be performed in arbitrary order. In case of AES, sixteen SubBytes operations need to be performed in every round. These substitution operations are independent of each other. Hence, they can be performed in arbitrary order. Shuffling of these operations means that during each execution of AES, randomly generated numbers are used to determine the sequence of the sixteen SubBytes operations. Shuffling randomizes the power consumption in a similar way as the random insertion of dummy operations. However, shuffling does not affect the throughput as much as the random insertion of dummy operations. In practice, shuffling and the random insertion of dummy operations are often combined. It should be noted that we have neither performed higher-order power analysis in this work, nor studied the resistance of the new implementation against it. So, we do not claim that this approach can counteract the higher-order DPA. It will remain as our future study.

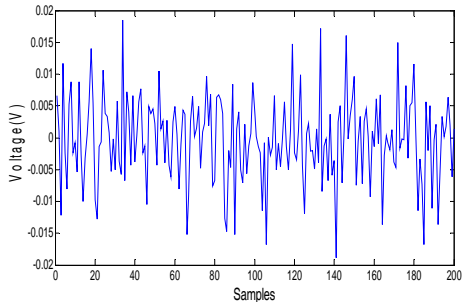

**Figure 13. The differential power traces for a correct subkey guess in the modified implementation.**

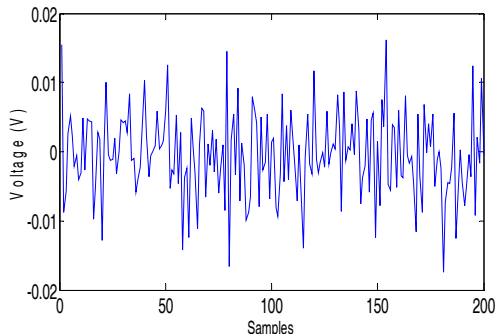

**Figure 14. The differential power traces for a wrong subkey guess in the modified implementation.**

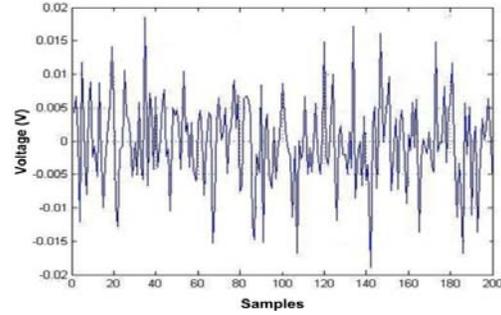

**Figure 15. The differential power traces for a known plaintext in the modified implementation with a specific set of parameters $\{\Phi, \lambda, \delta, \delta^{-1}\}$.**

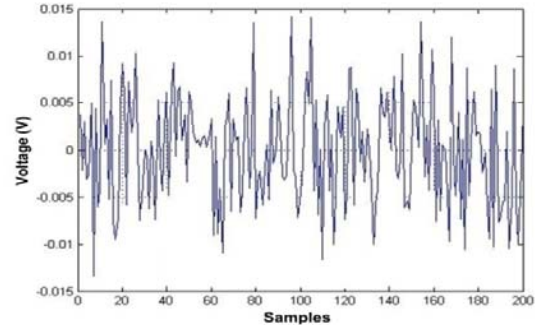

**Figure 16. The differential power traces for the same plaintext in the modified implementation with another set of parameters $\{\Phi, \lambda, \delta, \delta^{-1}\}$.**

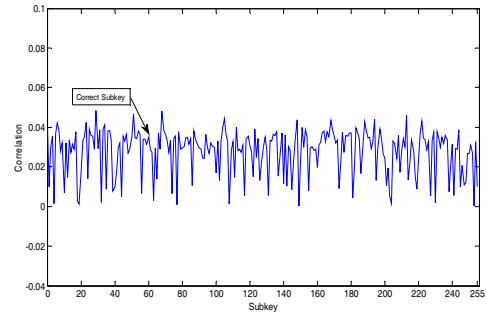

**Figure 17. Result of correlation analysis for recovering the correct subkey in the modified implementation.**

## 6. CONCLUSIONS

A novel AES implementation with a simple and integrated countermeasure against DPA was presented. The countermeasure is based on mathematical properties of Rijndael algorithm and retains perfect compatibility with the published Standard. The new design permits the construction of actual cores with very efficient area and speed characteristics, while still keeping a very high protection level. Although certain features of FPGAs make the practical implementation of power attacks considerably harder than in the smart card context, we have conducted relevant experimental tests and verified our idea practically. Nearly all the algorithms embedded in smart cards have been designed to resist at high level to linear, differential and high-order differential attacks, whereas nothing has been done to make them inherently resistant to DPA attacks. However, this work demonstrated that like with other attacks it is possible to design algorithms to be impervious to DPA, so when the next generation of cryptographic algorithms is designed they may be inherently secure against these attacks.